\documentclass[aps,prd,twocolumn,amsmath,amssymb]{revtex4}
\usepackage{natbib}
\usepackage{graphicx}
\usepackage{dcolumn}
\usepackage{bm}
\usepackage{multirow}

\begin{document} 

\title{Multiplicity spectra in $pp$ collisions at LHC energies in terms of Gamma and Tsallis distributions. }
\author{S. Sharma}
\author{M. Kaur}
\email{manjit@pu.ac.in} 

\affiliation{Department
of Physics, Panjab University, Chandigarh -160 014, India.}

\date{\today}

\begin{abstract}
In recent years the Tsallis statistics is gaining popularity in describing charged particle production and their properties, in particular $p_{T}$ spectra and the multiplicities in high energy particle collisions.~Motivated by its success, an analysis of the LHC data of proton-proton collisions at energies ranging from 0.9~TeV to 7~TeV in different rapidity windows for charged particle multiplicities has been done.~A comparative analysis is performed in terms of the Tsallis distribution, the Gamma distribution and the shifted-Gamma distribution.~An interesting observation on the inapplicability of these distributions at $\sqrt{s}$=7~TeV in the lower rapidity windows is intriguing.~The non-extensive nature of the Tsallis statistics is studied by determining the entropic index and its energy dependence.~The analysis is extrapolated to predict the multiplicity distribution at $\sqrt{s}$=14~TeV for one rapidity window, $|y| < 1.5$ with the Tsallis function.

\end{abstract}  

\maketitle

\section{INTRODUCTION}

In the high energy collisions of particles, several new particles are produced.~Experiments study the particle properties averaged over multiple collisions.~It is well known that properties such as the mean multiplicities, transverse momenta etc. follow rules of statistical mechanics.~Amongst the models of statistical particle production, the Tsallis distribution \cite{TS1,TS2} has been used extensively to describe transverse momentum spectra of particles stemming from high energy particle collisions.~Both for intermediate and high momenta regions, the particle production has been successfully described in the proton$-$proton~($pp$), antiproton$-$proton~($\overline{p}p$), nucleus$-$nucleus~($AA$) \cite{Biro,Urmo,Cley,Beck,Wilk} and $e^{+}e^{-}$ \cite{Bed,SS} collisions.~It has also been shown by K.~Urmossy et al \cite{Gam1} that in very high energy collisions, if only the momentum-energy conservation in hadronisation is taken into account, the average momentum distribution of produced particles can be considered as a micro-canonical generalisation of the Tsallis distribution.~Each final state particle created in a collision is identified with a microstate of a microcanonical esemble with scaling volume fluctuations.~If the momentum distribution in events with fixed multiplicity is micro-canonical, the shifted multiplicity has the Gamma-distribution.~In the Tsallis $q$-statistics, the entropy of standard statistical mechanics becomes non-extensive.~This non-extensive property of the entropy is then determined in terms of a parameter $q$, known as entropic index.~Which on account of its non-extensive behaviour exceeds unity.~Most of the analyses on the data from different kinds of collisions have been done to study the transverse momentum distributions and fragmentation functions.~At LHC energies, some analyses \cite{DE, PL,T,Azmi} have been done to study the $p_{T}$ spectra by using the Tsallis distribution.~However, the analysis of the multiplicities has been done only in very few cases \cite{Jan1, Jan2, Urmo3} using different approaches.~In addition, analyses of the data from various experiments at the RHIC and at the LHC have shown excellent fits to the transverse momentum distributions with the Tsallis-like distribution \cite{JCley}.~A. Capella et al \cite{Cap} have studied the multiplicities at the LHC in the Pomeron model.

In this paper, the first study of multiplicity distributions is reported on the proton-proton collisions at the LHC energies in the restricted central rapidity windows.~Energy-momentum conservation strongly influences the multiplicity distribution for the full phase space.~The distribution in restricted rapidity windows however, is less prone to such constraints and thus can be expected to be a more sensitive probe to the underlying dynamics of QCD, as inferred in references \cite{Gam1, Gam2}.~We also study the dependence of the entropic index $q$, an important parameter in the Tsallis $q$-statistics, on energy of the
collisions.

After this brief introduction in section I, we describe the essential steps of the distributions, Tsallis, Gamma and shifted-Gamma and the basic definition of rapidity used, in section II.~Section III gives details of the data used and results from the comparison of our analysis of the three distributions.~Section IV presents the conclusions.
   
\section{PARTICLE PRODUCTION AND THE DISTRIBUTION}
The charged particles produced in a collision are emitted at all angles and measured in terms of rapidity defined as $y = -ln(\frac{E+p_{L}}{E-p_{L}})$ where E is the particle energy and $p_{L}$ is the longitudinal momentum.~The number of particles produced are distributed according to some probability distribution function (PDF) with mean of the distribution coinciding with the average number, called the average multiplicity.~We discuss three such PDFs in the following section;

\subsection{The Gamma and the shifted-Gamma distributions}
 
The Gamma is a very basic distribution which describes the multiplicity distributions at lower energies very well.~Inclusive data $e^{+}e^{-}\rightarrow h^{\pm} + X$ from LEP experiments were studied by Urmossy et al \cite{Gam1} by considering a sample of two$-$jet events.~They used the Boltzman-Gibbs and micro-canonical distributions in one dimension.~They showed that the Gamma distribution of the shifted multiplicity, N-N0 can result in a Tsallis or micro-canonical Tsallis shaped spectrum.  

The probability density function for the Gamma and the shifted-Gamma distributions are given below;
\begin{equation}
P_N = AN^{\alpha-1} exp^{-\beta{N}} \hspace{0.8cm} 
\end{equation}  
with $\alpha$ the scale parameter, $\beta$ the shape parameter and $A$ are the fit parameters of the distribution.~The average momentum distribution is Tsallis. 

A shift in the multiplicity $N \rightarrow (N-N0 )$ is exploited,
without violating the KNO scaling \cite{KNO}, then the averaging is done over the multiplicity distribution;
\begin{equation}
P_N = A(N-N0)^{\alpha^{\prime}-1} exp^{-\beta^{\prime}({N-N0})} \hspace{0.8cm} 
\end{equation}
The resulting momentum distribution is a possible micro-canonical generalisation of the Tsallis.
~The shift in the multiplicity has been chosen to be $N0 = 1 + 2/D$ where D is the dimensionality of phase space.~The details can be found in \cite{Gam1, Gam2}

\subsection{The Tsalls distribution}
The Tsallis q-statistics deals with entropy in the usual Boltzman-Gibbs thermo-statistics modified by introducing $q$-parameter. 
~For a given thermo-dynamical system, when divided into two subsystems, the Tsallis entropy no longer remains extensive, but is defined as
 \begin{equation}
 S_q(A,B)= S_A + S_B+ (1-q)S_{A}S_{B}
 \end{equation}
where $q$ is known as the entropic index with value $q>1$ and $(1-q)$ measures the departure of entropy from its extensive behaviour.
Assuming the interaction as a canonical ensemble of N particles, the partition function is defined through the probability as  
\begin{equation}
P_N = \frac{Z^{N}_q}{Z}
\end{equation}
where Z represents  the total partition function and $Z^{N}_q$ represents partition function at a particular multiplicity.~C.E. Aguiar et al \cite{TS2} have discussed in detail the method for calculating N particles partition function and deriving the probability distribution.~Details of these calculations can be obtained from this reference.
  
\section{RESULTS}

The experimental data of proton-proton collisions at the Large Hadron Collider (LHC) obtained by the CMS experiment for different energies are analysed.~The data analysed are at $\sqrt{s}$ = 0.9~TeV, 2.34~TeV and 7~TeV in the restricted rapidity windows of $|y|<$ 0.5, 1.0, 1.5, 2.0 and 2.4.~The experimental data \cite{CMS} are fitted with the distributions from the Tsallis $q$-statistics, the Gamma distribution and the shifted-Gamma distribution.~The results are discussed in the following sections;
 
\subsection{ The Gamma vs. the Tsallis  distribution}
The probabilities from the Gamma distribution, the shifted-Gamma distribution and the Tsallis distribution are calculated using equations (1), (2) and (4).~Fits to the data are shown in figures~1-3.~Table~I gives the parameters of the fits for all the rapidity windows at energies 0.9~TeV 2.36~TeV and 7~TeV respectively and a comparison of corresponding $\chi^{2}/ndf$ and $p$-values are given in the table~II.~While fitting we consider the probability distribution for 7~TeV extending up to the continuous range of $N$ values.~Beyond this the statistics is very low and the probability falls below .001 leading to the fit parameters with very large errors, particularly for the Tsallis distribution.     
  
In order to study the behaviour of the three distributions for multiplicities of charged particles produced with higher transverse momenta $p_{T}$, the analysis is extended to the data $\sqrt{s}$ = 0.9~TeV, 2.36~TeV and 7~TeV in the restricted rapidity window of $|y|<$ 2.4 and $p_{T} >$ 500~MeV.~Figure~4 shows the results of fits for the three distributions.~The fit parameters are given in the table~III and the $\chi^{2}/ndf$ and $p$-values in the table~IV. 

One finds that both the Tsallis and the shifted-Gamma distributions reproduce the data very well in most of the rapidity windows at the three energies in comparison to the Gamma distribution.~However all the distributions fail for rapidity windows $|y|<$ 0.5 and $|y|<$ 1.0 with $p$-values corresponding to $CL< 0.10\%$.~The detailed comparison between the three functions is shown in the table II where $\chi^{2}/ndf$ and $p$-values at all energies for all rapidity windows are compared. 
It is found that the $\chi^{2}/ndf$ values are  comparable for the Tsallis and the shifted-Gamma fits with $p$-values corresponding to $CL> 0.1\%$ as compared to the Gamma fits.~This is true for nearly all the rapidity windows at all the energies.
 
Results from the table~III and IV for the fits of the three distributions to the data with $p_{T} > $500~MeV in the $|y|<$ 2.4 rapidity window, show that for 7~TeV data $p$-values correspond to $CL<0.1\%$ and hence all the fits are statistically excluded at this energy.~However at the other two energies, all fits are able to reproduce the data, with the shifted-Gamma giving the best.

\begin{figure}
\includegraphics[width=3.8 in , height= 2.9 in]{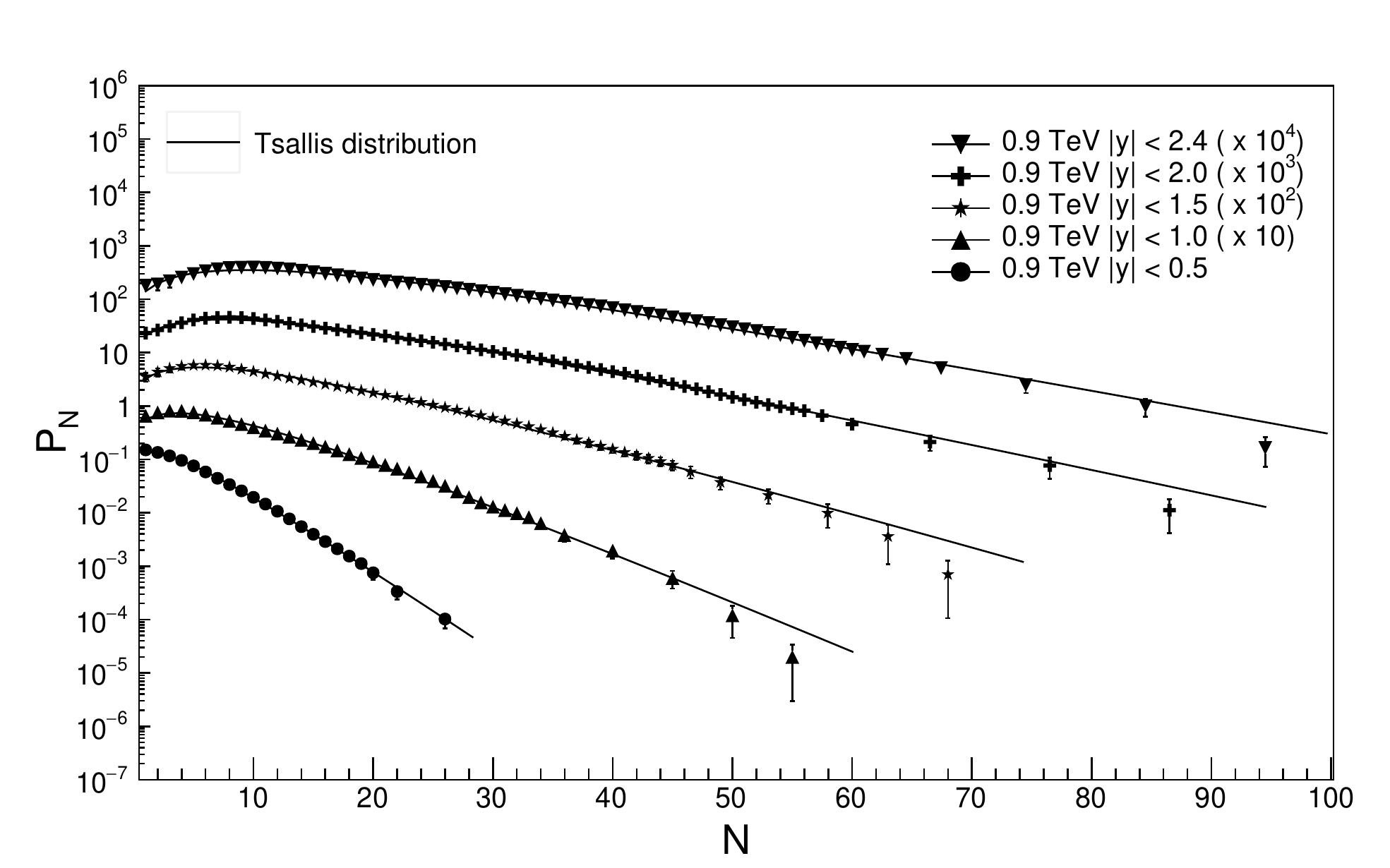}
\includegraphics[width=3.8 in , height= 2.9 in]{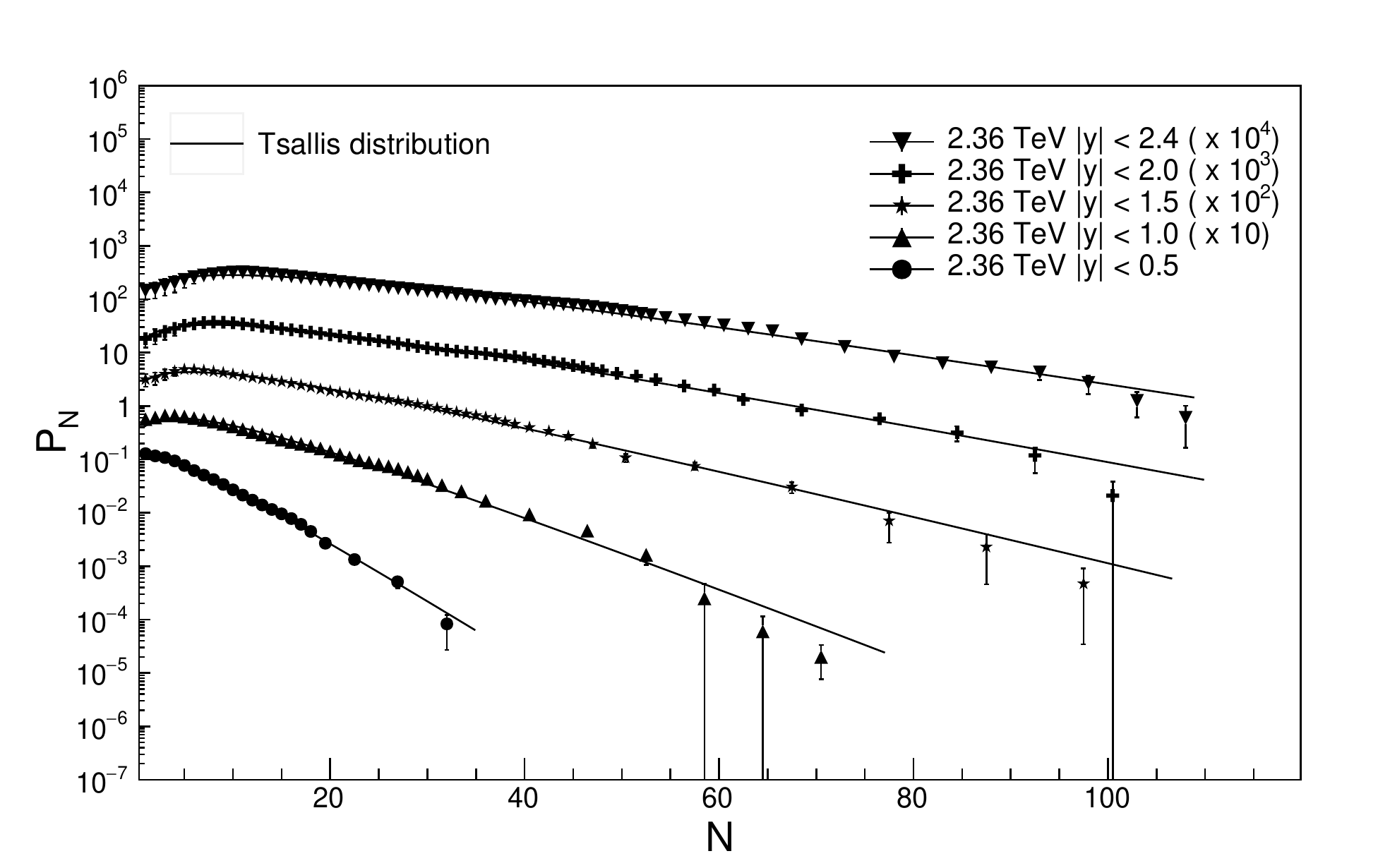}
\includegraphics[width=3.8 in , height= 2.9 in]{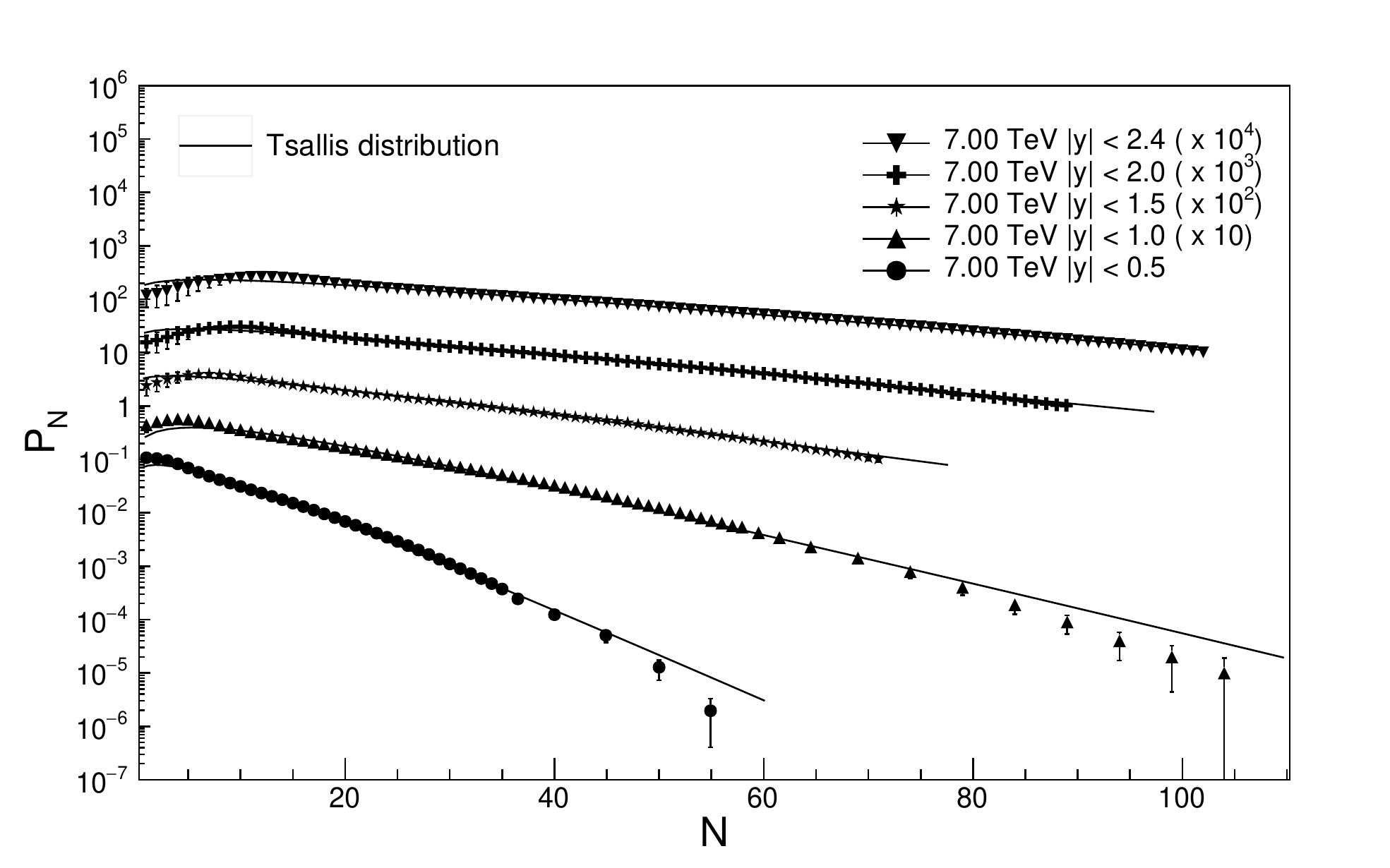}

\caption{The charged particle multiplicity distributions measured in $pp$ collisions by the CMS experiment with the fits by the Tsallis distribution.~Each successive distribution is multiplied by a factor of 10.}
\end{figure}

\begin{figure}

\includegraphics[width= 3.8 in , height= 2.9 in]{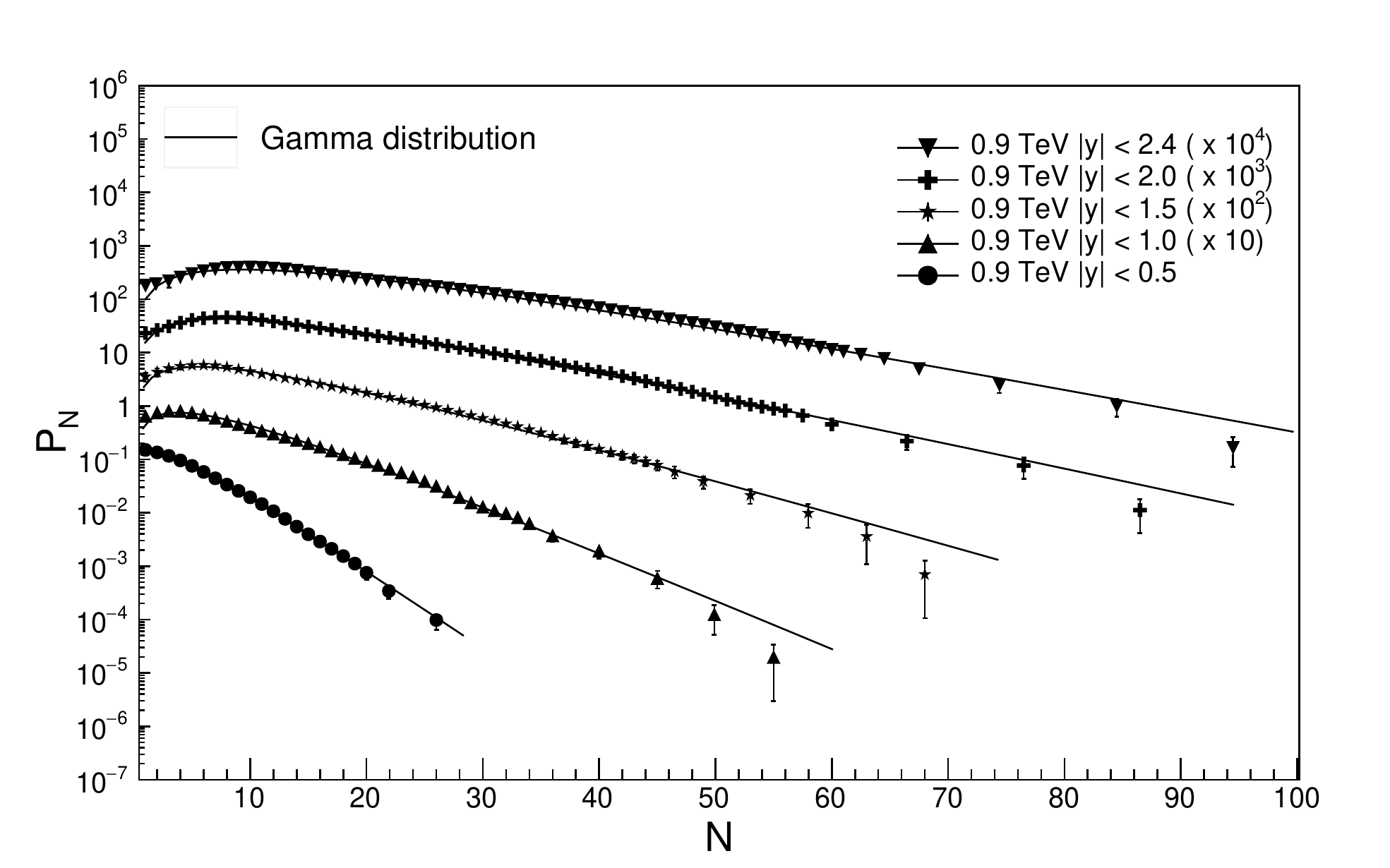}
\includegraphics[width= 3.8 in , height= 2.9 in]{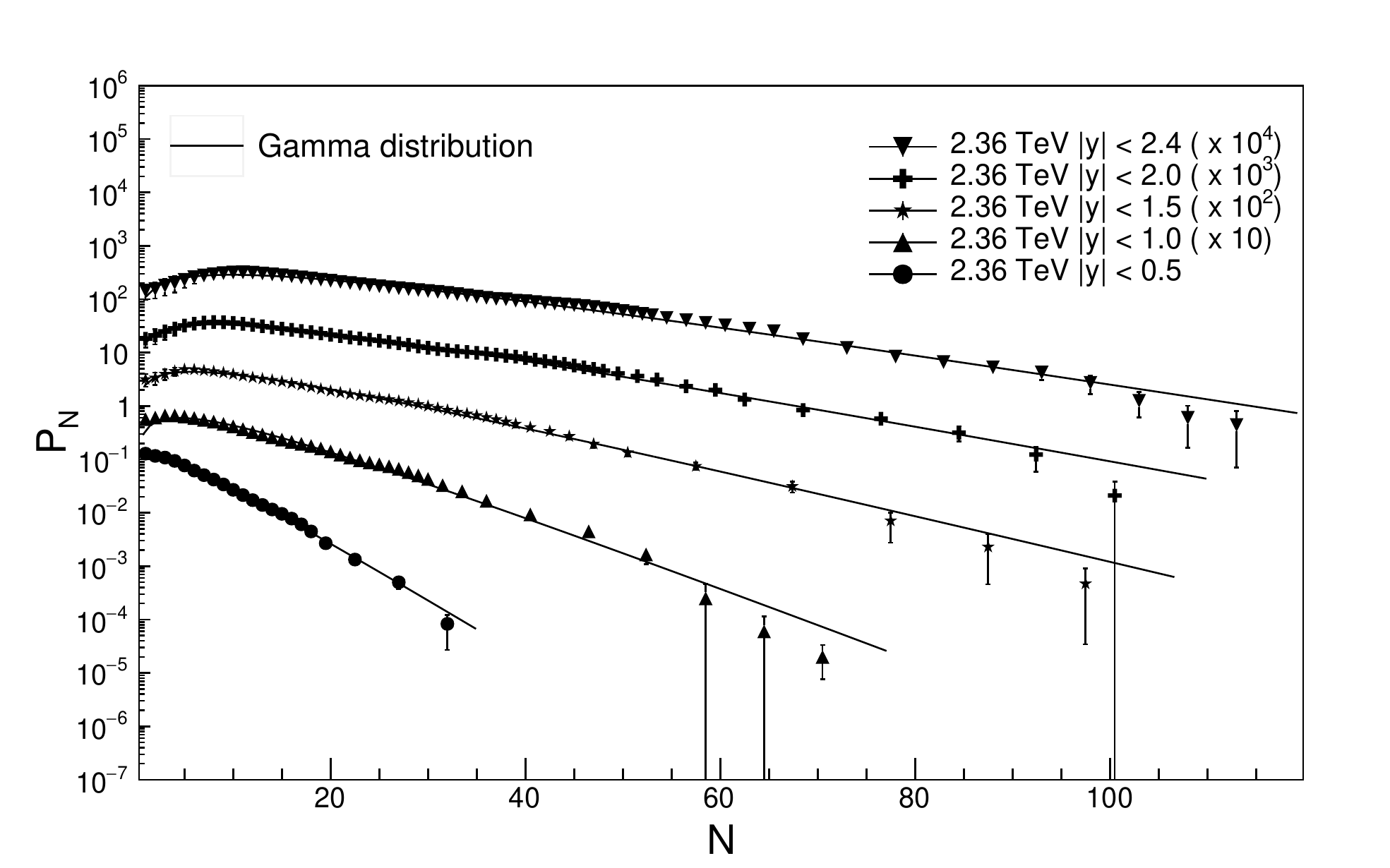}
\includegraphics[width=3.8 in , height= 2.9 in]{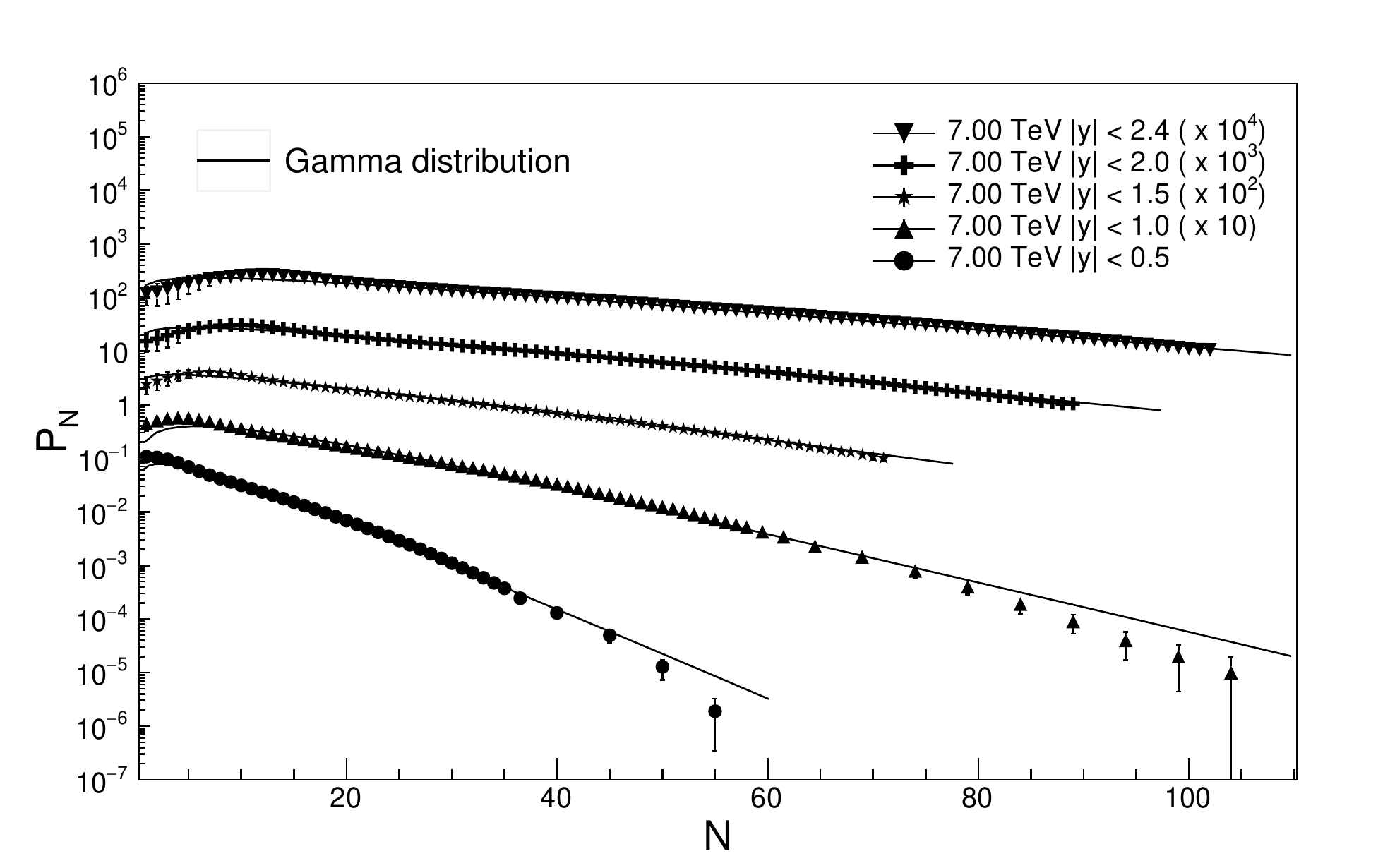}

\caption{The charged particle multiplicity distributions measured in $pp$ collisions by the CMS experiment with the fits by the Gamma distribution.~Each successive distribution is multiplied by a factor of 10.}
\end{figure}

\begin{figure}
\includegraphics[width= 3.8 in , height= 2.9 in]{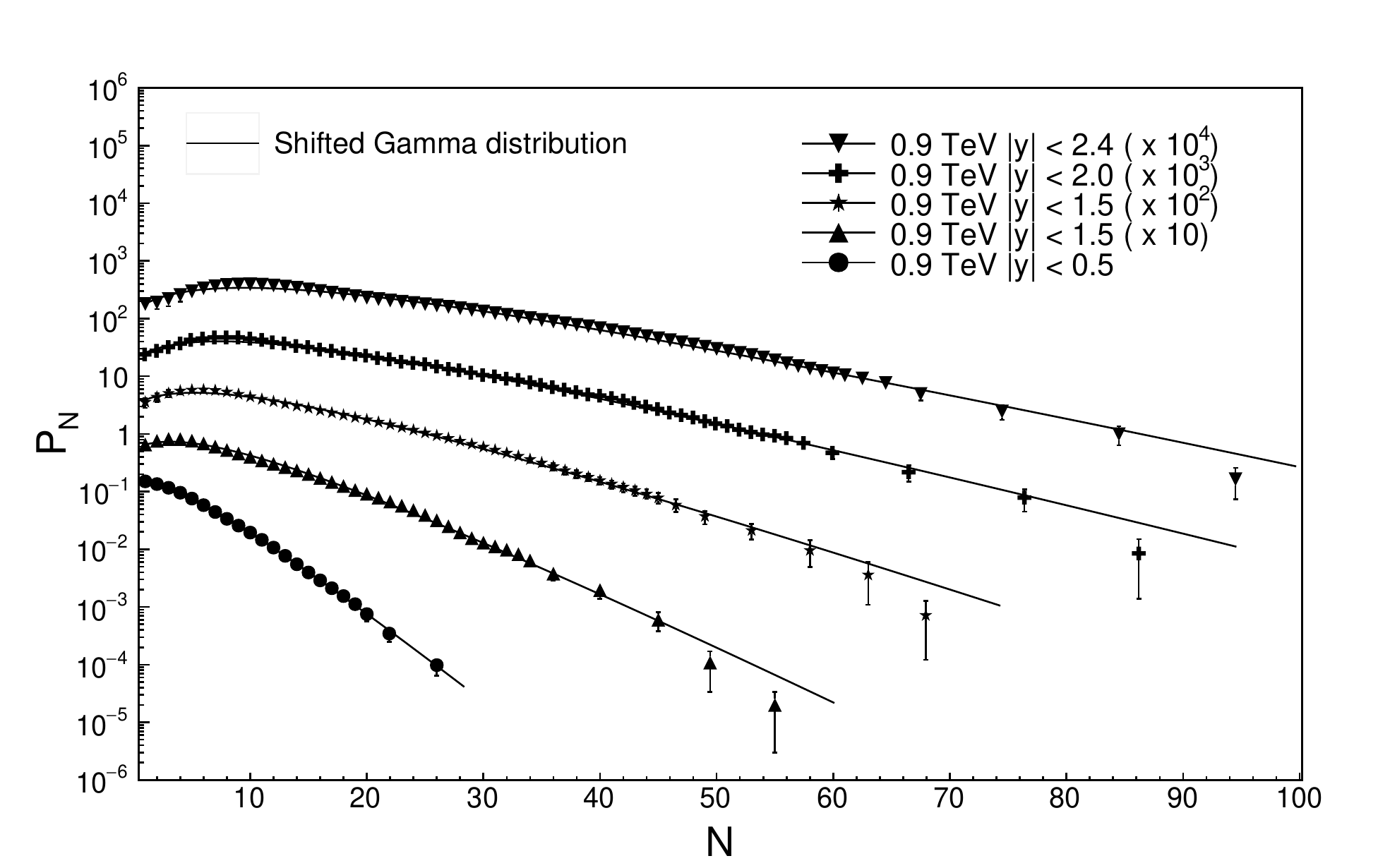}
\includegraphics[width= 3.8 in , height= 2.9 in]{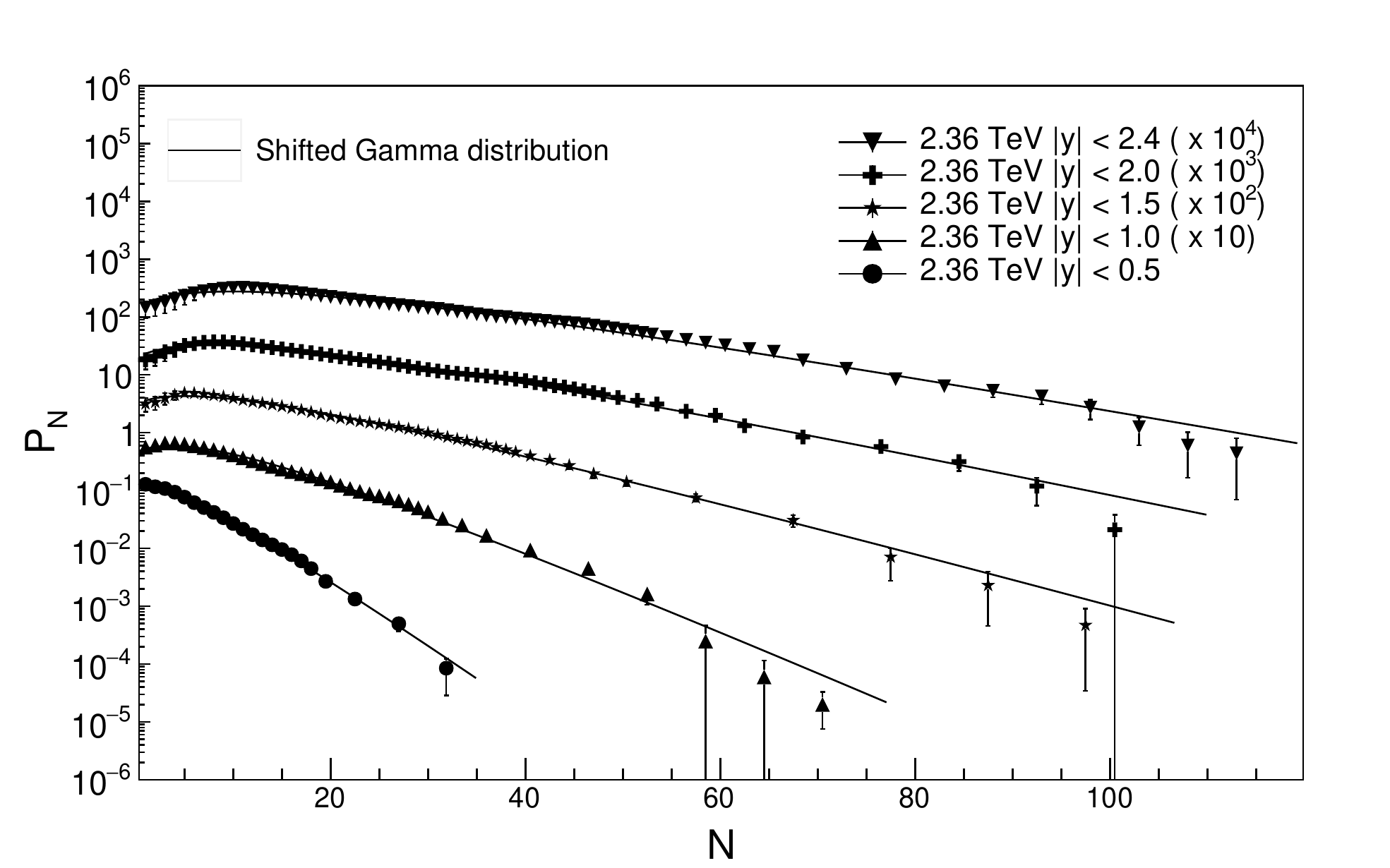}
\includegraphics[width=3.8 in , height= 2.9 in]{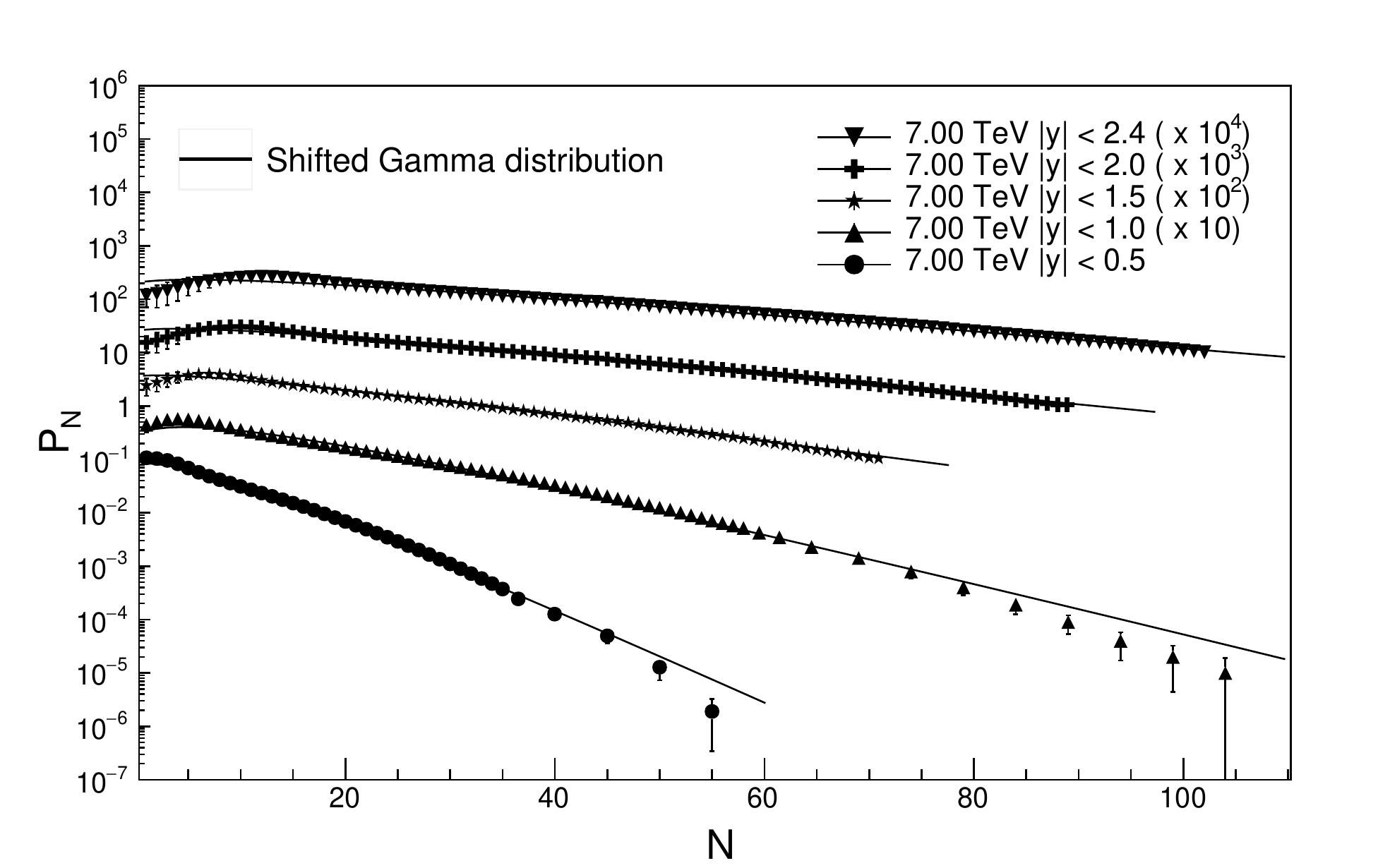}

\caption{The charged particle multiplicity distributions measured in $pp$ collisions by the CMS experiment with the fits by the shifted-Gamma distribution.~Each successive distribution is multiplied by a factor of 10.}
\end{figure} 

\begin{figure}
\includegraphics[width= 3.8 in , height= 2.9 in]{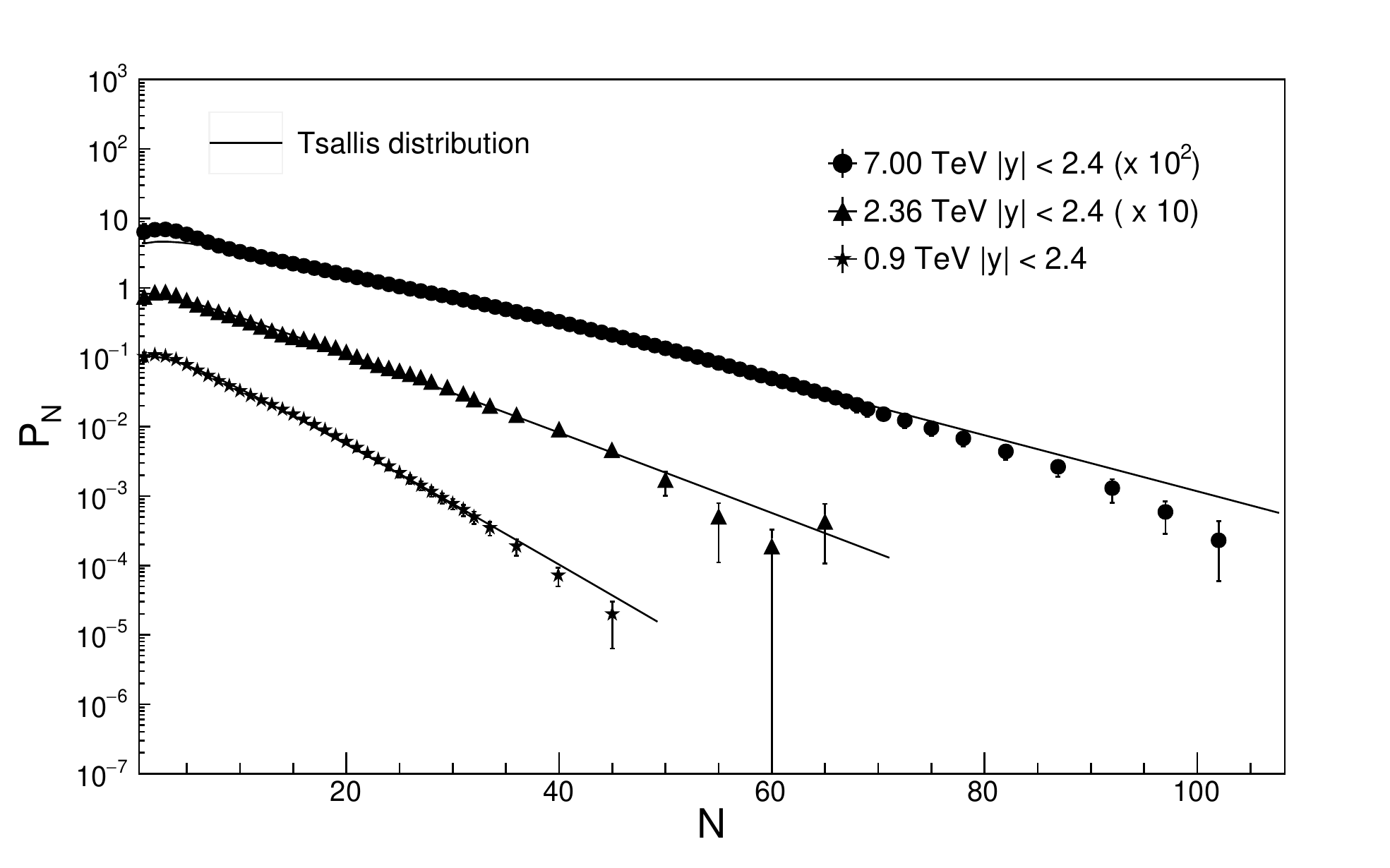}
\includegraphics[width= 3.8 in , height= 2.9 in]{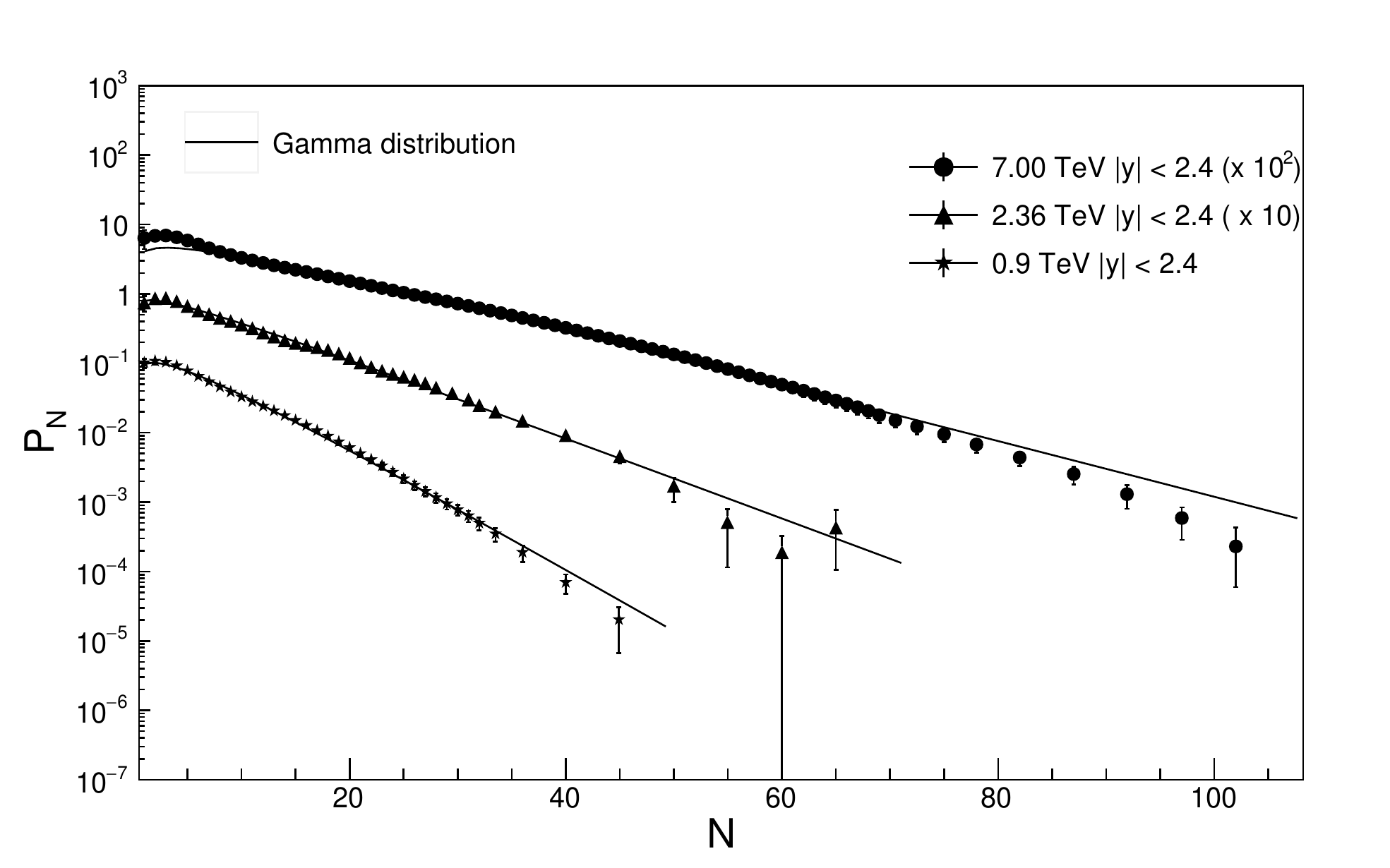}
\includegraphics[width=3.8 in , height= 2.9 in]{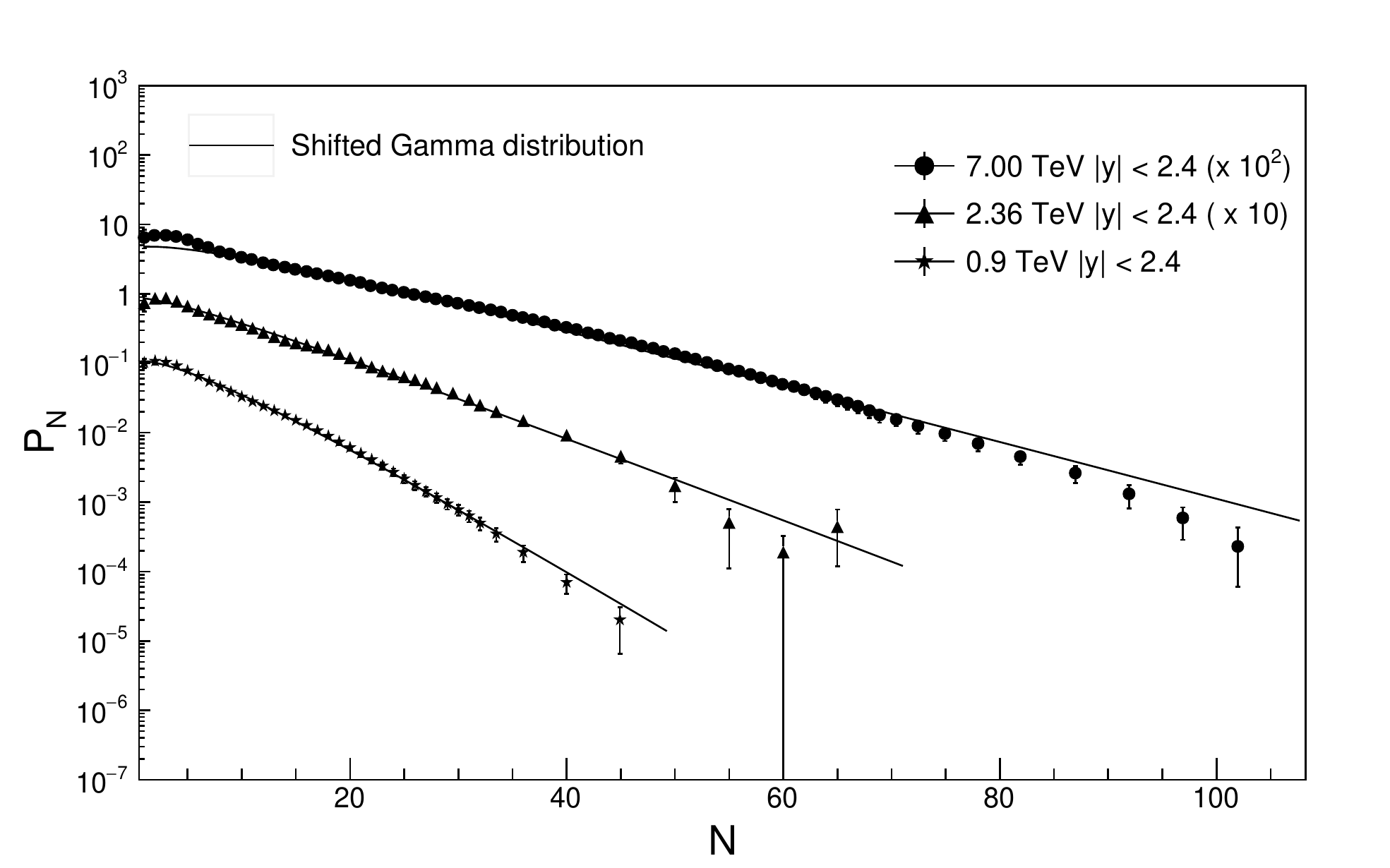}

\caption{The charged multiplicity distributions for particles with $|y|<2.4$, $p_{T} >$ 500~MeV measured by the CMS experiment with the fits by the Tsallis, the Gamma and the shifted-Gamma distributions.~Each successive distribution is multiplied by a factor of 10.}
\end{figure} 

From the tables I \& II, it is also observed that for the Gamma distribution, $\beta$ values decrease with energy as well as with rapidity.~This is as per the expected trend.~Similarly for the shifted-Gamma distribution, $\beta^{\prime}$ values decrease with energy as well as with rapidity.~Both $\beta$ and $\beta^{\prime}$ measure the shape parameters for the two distributions.~For the Tsallis distribution, the $q$ value which measures the entropic index of the Tsallis statistics, increases with energy and exceeds unity in each case.~This confirms that the Tsallis statistics becomes non-extensive.~The parameter $K$ also determines the shape of the distribution, it becomes binomial-like if $K$ becomes negative.

Figure~5, shows energy dependence of the entropic parameter of the Tsallis function in various rapidity windows.~The dependence can be parameterised as a power law, $q=A\sqrt{s}^B$.~The fit parameters $A$ and $B$ are listed in the table~V.~It may be observed that the value of $B$ is constant and independent of rapidity.~In a study of the sytematic properties of Tsallis distribution, J. Cleyman et al \cite{JCley} have studied the energy dependence of parameters of Tsallis distribution in $pp$ collisions and shown that $q$ has a weak dependence on beam energy.~From the transverse momentum distributions, they have determined the $q$ values from the ATLAS data \cite{Aad} at $\sqrt{s}$=0.9, 2.36 and 7~TeV  as 1.1217$\pm$0.0007, 1.1419$\pm$0.0025 and 1.1479$\pm$0.0008.~These values agree very closely with the values we obtain from multiplicity distribution fits in rapidity $|y|<$ 2.4 for the CMS data; 1.055$\pm$0.008, 1.136$\pm$0.031 and 1.228$\pm$0.055.~Small differences are expected from the slightly different phase spaces considered in the two cases.
Figure~6 shows the multiplicity distribution of charged particles predicted in the $pp$ collisions in the restricted rapidity window of $|y|<1.5$ at $\sqrt{s}$=14~TeV.~The value of $q$ is predicted as 1.476$\pm$0.108.~Similar predictions can be made for other higher rapidity windows.

In a further investigation of the failure of all distributions at 7~TeV, we consider 2-component approach, soft and semi-hard component-structure in the multi-particle production.~This leads to the division of the distribution in terms of soft events(events without minijets) and the semi-hard events(events with minijets).~A distribution is then produced as a weighted superposition of the two components, the weight $\alpha_{soft}$ being the fraction of soft events, as below;
\small
\small
\begin{equation}
P(n)=\alpha_{soft}P_{soft}^{MD}(n)+(1-\alpha_{soft})P_{semi-hard}^{MD}(n)
\end{equation} 
\normalsize
The multiplicity distribution (MD) of each component being one of the three distributions under consideration; Gamma, shifted-Gamma or Tsallis distribution.~The idea of this superposition was first suggested by C. Fuglesang \cite{Fuse} in order to explain the negative binomial regularity violations.~The concept originates from purely phenomenological and very simple considerations.~The two fragments of the distribution suggests the presence of the substructure.~For example, by using this approach, fits of the data at 7~TeV with the the shifted-Gamma distribution, in the rapidity windows $|y|<$ 0.5, $|y|<$ 1.0 and $|y|<$ 2.4 reduces the $\chi^{2}/ndf$ values by a large factor and the distributions become statistically significant with $CL>0.1\%$.~The results are shown in the table~VI.~Since the $\alpha_{soft}$ value is not available, it was input in the distribution and iterated to obtain the best fit.~This observation indicates that at higher energies, the contribution of the events with mini jets grows.~Similar fits when used for data at other energies and rapidity regions also reduce the $\chi^{2}/ndf$ in every case.~However, in case of Tsallis distribution, the number of parameters becomes as large as 9 and the fit values of the parameters have very large errors.  
\begin{figure}[h]
\includegraphics[width=3.7 in]{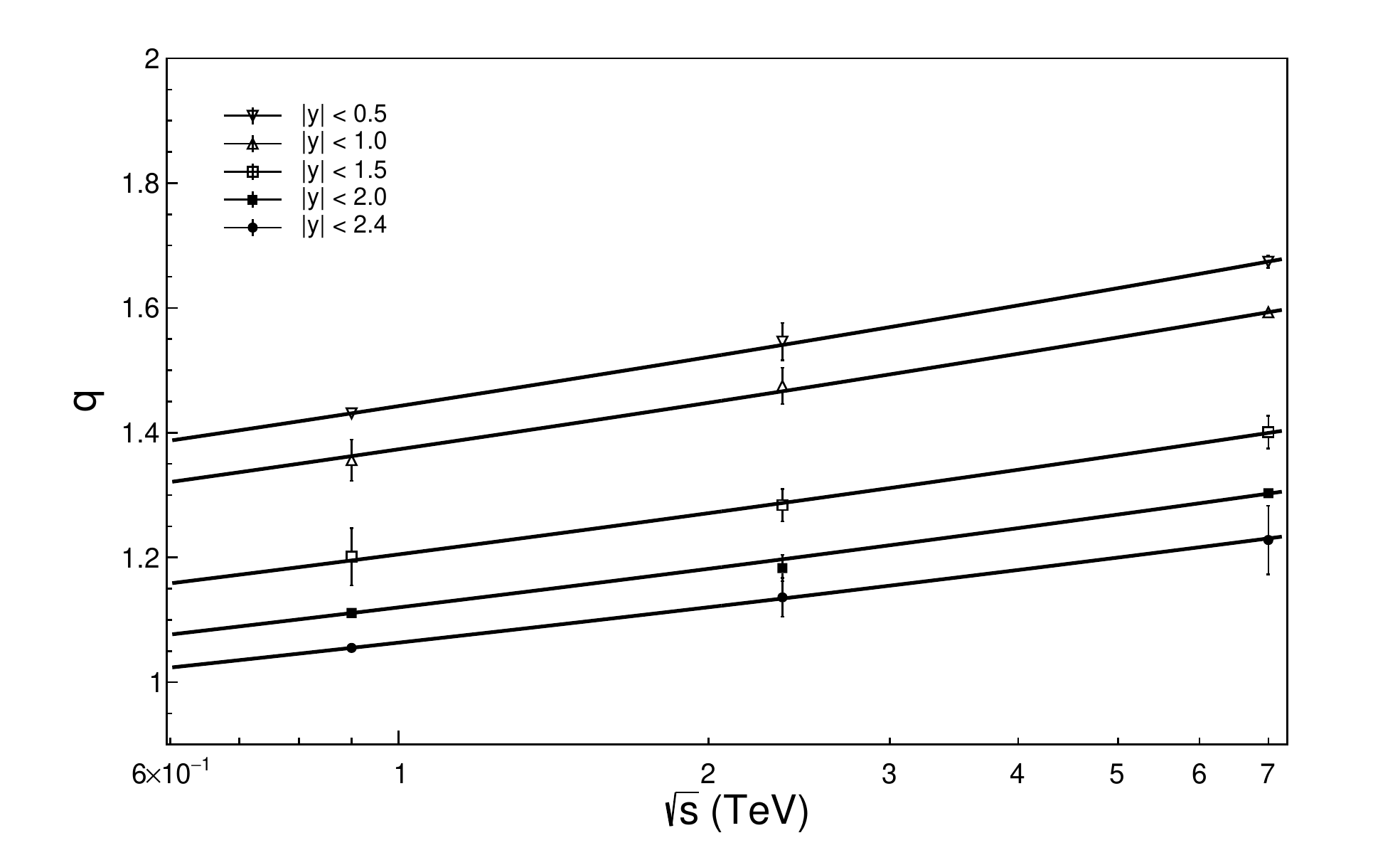}
\caption{The energy dependence of the non-extensive entropic parameter of the Tsallis function in different rapidity windows.}
\end{figure}

\begin{figure}[h]
\includegraphics[width=3.9 in]{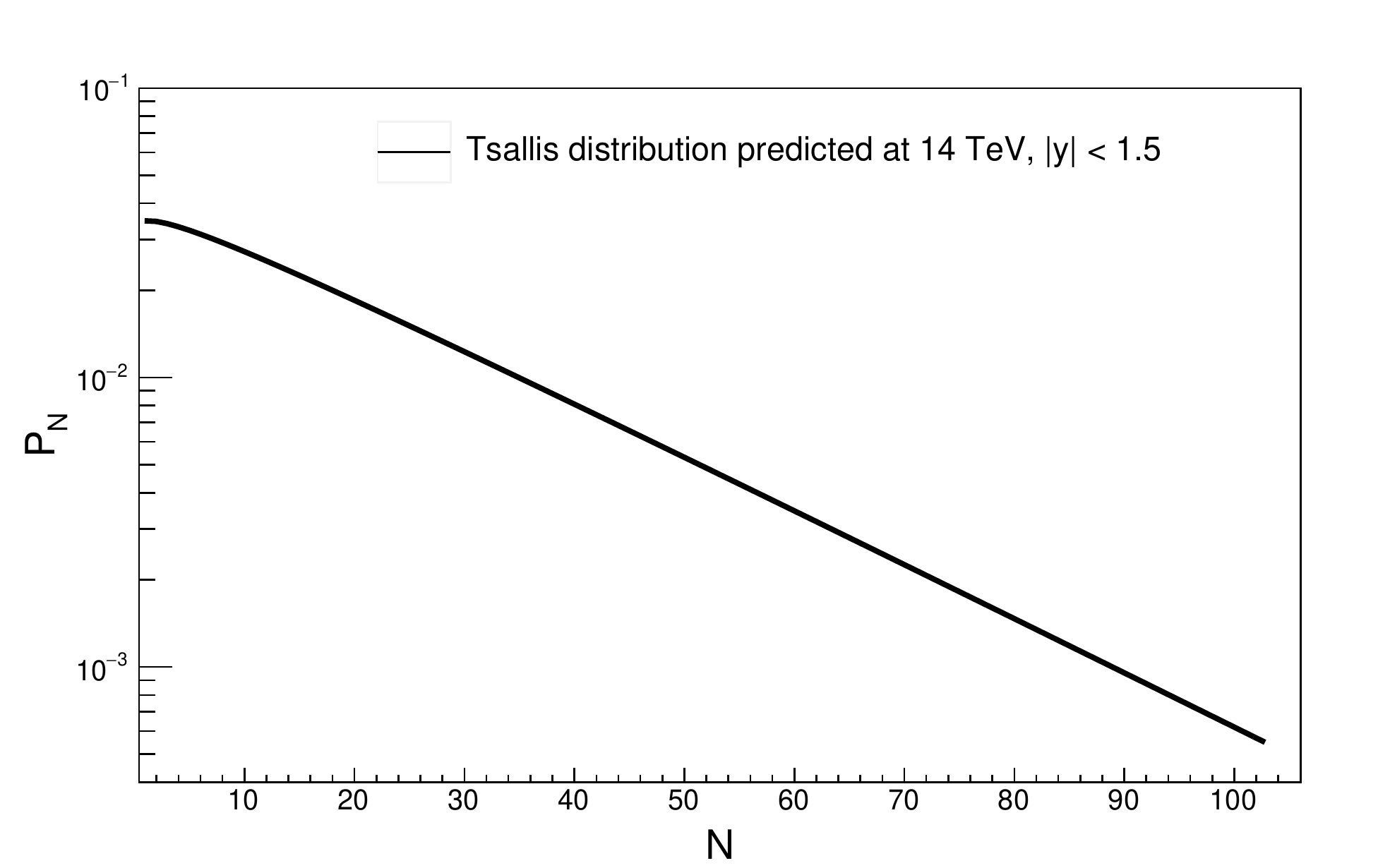}
\caption{The multiplicity distribution predicted for $pp$ collisions at $\sqrt{s}$=14~TeV in $|y| < 1.5$ window.}
\end{figure}

\section{CONCLUSION}
The data on charged multiplicities at the LHC energies, $\sqrt{s}$=0.9, 2.36 and 7~TeV in proton-proton collisions in the restricted rapidity windows obtained by CMS experiment have been used for a detailed analysis of the multiplicities.~Analysis has also been done for the particles emitted with $p_{T} >$500~MeV in the largest rapidity window $|y|< 2.4$.~A comparison of the Gamma distribution, the shifted-Gamma distribution and the Tsallis distribution has been done.~The relevance of the comparison is due to the similar nature of the three distributions.~When the multiplicity has Gamma distribution, the average momentum distribution of particles is Tsallis like.~For the shifted-Gamma distribution, the avearge momentum distribution is a possible micro-canonical generalisation of the Tsallis.~A comparison of the $\chi^{2}/ndf$ and p-values in table II, shows that all the three distributions reproduce the data in most of the rapidity windows at 0.9~TeV and 2.36~TeV.~However at 7~TeV, all distributions fail and are statistically excluded with $CL<0.1\%$ in the two lower rapidity windows, $|y|< 0.5$ and $|y|< 1.0$.Interestingly, for the particles emitted with $p_{T} >$500~MeV, at $\sqrt{s}$=7~TeV, all the distributions fail with $CL<0.1\%$.~This indicates the possible dynamical changes in the particle production at such high collision energy.
  
~Overall the Tsallis fits and the shifted-Gamma fits are comparable and much better than the Gamma fits.~The value of $q$ determined at each center of mass energy and in different rapidity windows exceeds unity.~The value of $q$ decreases with the increase in rapidity window size at a given energy.~For collisions in the same size rapidity window at different energies, the $q$ value increases with energy as shown by values in tables~I and III, indicating that for collisions at higher energies, the non-extensive behaviour of entropy becomes more pronounced.~The energy dependence of $q$ is described by the power law.~The parametrisation as a power law is inspired by the observation that single particle energy distribution obeys a power law behaviour \cite{Gaz}.~Entropy being determined by the energy fluctuations, influences the $q$ values.~The $q$ values obtained from our analysis agree very well with the results from analysis of the data from the ATLAS experiment.~Thus the consistent results from two different analyses confirm the $q$-values being weakly dependent on the beam energy, non-extensive nature of
the entropy of collisions and a change in dynamics of particle production at high energies.\\

\begin{center}
	{\bf ACKNOWLEDGMENT}
\end{center}	
S.S. is grateful to the Department of Science and Technology, Government of India for the research fellowship grant.
  
\newpage


\begin{table*}[t]
\begin{tabular}{|c|c|c|c|c|c|c|c|c|c|}
\hline

 &\multicolumn{2}{|c|}{}& \multicolumn{2}{|c|}{} &\multicolumn{4}{|c|}{}   \\
 Rapidity  & \multicolumn{2}{|c|}{Gamma distribution } & \multicolumn{2}{|c|}{shifted-Gamma distribution} &\multicolumn{4}{|c|}{Tsallis distribution}\\ 
 Interval  & \multicolumn{2}{|c|}{} & \multicolumn{2}{|c|}{} &\multicolumn{4}{|c|}{}\\ \hline
 
  	  		\multicolumn{9}{|c|}{}  \\ 
  			 \multicolumn{9}{|c|}{$\sqrt{s}$ = 0.9 TeV}  \\       
			  \multicolumn{9}{|c|}{}    \\\hline
		&	          &			 &	  &	  &   &   &  &  \\	  
  	  $|y|$ & $\alpha$  & $\beta$ &$\alpha'$ & $\beta'$ &  $nV$ & $nv_{0}$ & K & q \\ 
  	 	
  	 	  &	          &			 &	  &	  &   &   &  &  \\\hline 
  	 
0.5	&1.515 $\pm$ 0.068	&0.352 $\pm$ 0.010 &2.265 $\pm$ 0.156 &	0.395 $\pm$ 0.015& 1.932 $\pm$ 0.152	&0.286 $\pm$ 0.135	&1.726 $\pm$ 0.098	&1.431 $\pm$ 0.004\\\hline

1.0	&1.848 $\pm$ 0.061	&0.223 $\pm$ 0.004& 2.498 $\pm$ 0.098 &	0.244 $\pm$ 0.005	& 3.043 $\pm$ 0.091	&0.317 $\pm$ 0.019	&2.084 $\pm$ 0.080	&1.356 $\pm$ 0.033\\\hline

1.5	&1.844 $\pm$ 0.058	&0.153 $\pm$ 0.003	& 2.371 $\pm$ 0.081 &	0.167 $\pm$ 0.004 & 4.067 $\pm$ 0.661	&0.134 $\pm$ 0.190	&2.028 $\pm$ 0.070	&1.201 $\pm$ 0.046\\\hline

2.0	&1.865 $\pm$ 0.053	&0.117 $\pm$ 0.002	& 2.331 $\pm$ 0.067&	0.128 $\pm$ 0.002 & 4.338 $\pm$ 0.120	&0.375 $\pm$ 0.078	&2.025 $\pm$ 0.055	&1.111 $\pm$ 0.003 \\\hline

2.4	&1.944 $\pm$ 0.054	&0.102 $\pm$ 0.002	& 2.382 $\pm$ 0.062 &	0.111 $\pm$ 0.002 & 4.802 $\pm$ 0.071	&0.417 $\pm$ 0.010	&2.101 $\pm$ 0.058	&1.055 $\pm$ 0.008 \\\hline

 	\multicolumn{9}{|c|}{}  \\ 
  			 \multicolumn{9}{|c|}{$\sqrt{s}$ = 2.36 TeV}  \\       
			  \multicolumn{9}{|c|}{}    \\\hline

0.5	&1.379 $\pm$ 0.068	&0.260 $\pm$ 0.009 & 1.855 $\pm$ 0.141 &	0.284 $\pm$ 0.012 &2.271 $\pm$ 0.163	&0.150 $\pm$ 0.180	&1.501 $\pm$ 0.088	&1.546 $\pm$ 0.030 \\\hline

1.0	&1.743 $\pm$ 0.060	&0.168 $\pm$ 0.003 & 2.187 $\pm$ 0.089 &	0.179 $\pm$ 0.004 & 3.702 $\pm$ 0.102	&0.152 $\pm$ 0.011	&1.892 $\pm$ 0.073	&1.475 $\pm$ 0.029\\\hline

1.5	&1.595 $\pm$ 0.057	&0.105 $\pm$ 0.003 & 1.892 $\pm$ 0.081 &	0.112 $\pm$ 0.003 & 4.186 $\pm$ 0.426	&0.257 $\pm$ 0.251	&1.682 $\pm$ 0.066	&1.284 $\pm$ 0.026\\\hline

2.0	&1.644 $\pm$ 0.057	&0.082 $\pm$ 0.002 & 1.931 $\pm$ 0.075 &	0.087 $\pm$ 0.002 & 4.770 $\pm$ 0.350	&0.431 $\pm$ 0.305	&1.731 $\pm$ 0.063	&1.183 $\pm$ 0.021\\\hline

2.4	&1.684 $\pm$ 0.057	&0.070 $\pm$ 0.002 & 1.959 $\pm$ 0.071 &	0.075 $\pm$ 0.002 & 5.227 $\pm$ 0.394	&0.441 $\pm$ 0.271	&1.751 $\pm$ 0.059	&1.136 $\pm$ 0.031\\\hline

  		\multicolumn{9}{|c|}{}  \\ 
  			 \multicolumn{9}{|c|}{$\sqrt{s}$ = 7.00 TeV}  \\       
			  \multicolumn{9}{|c|}{}    \\\hline
		
0.5	&1.461 $\pm$ 0.050	&0.201 $\pm$ 0.007 & 1.841 $\pm$ 0.083 &	0.214 $\pm$ 0.004	&2.871 $\pm$ 0.068	&0.136 $\pm$ 0.005	&1.571 $\pm$ 0.062	&1.674 $\pm$ 0.010\\\hline	
1.0	&1.645 $\pm$ 0.044	&0.113 $\pm$ 0.003 & 1.897 $\pm$ 0.059 &	0.118 $\pm$ 0.002	&3.745 $\pm$ 0.103	&0.159 $\pm$ 0.014	&1.725 $\pm$ 0.043	&1.593 $\pm$ 0.004\\\hline	
1.5	&1.767 $\pm$ 0.041	&0.081 $\pm$ 0.001 & 1.317 $\pm$ 0.059 &	0.063 $\pm$ 0.002	&4.232 $\pm$ 0.064	&0.443 $\pm$ 0.296	&1.247 $\pm$ 0.035	&1.401 $\pm$ 0.026\\\hline	
2.0	&1.738 $\pm$ 0.039	&0.061 $\pm$ 0.002	& 1.325 $\pm$ 0.050 &	0.048 $\pm$ 0.001 &5.101 $\pm$ 0.122	&0.334 $\pm$ 0.026	&1.265 $\pm$ 0.040	&1.303 $\pm$ 0.007\\\hline	
2.4	&1.506 $\pm$ 0.031	&0.046 $\pm$ 0.001 & 1.344 $\pm$ 0.047 &	0.040 $\pm$ 0.001	&5.769 $\pm$ 0.217 &	0.221 $\pm$ 0.218	& 1.284 $\pm$ 0.035	& 1.228 $\pm$ 0.055\\\hline

\end{tabular}
  \caption{Fit parameters of the distributions for all rapidity windows for the $pp$ data.}
\end{table*}

\begin{table*}[t]
\begin{tabular}{|c|c|c|c|c|c|c|c|}
\hline

 &  &\multicolumn{2}{|c|}{} & \multicolumn{2}{|c|}{} &\multicolumn{2}{|c|}{}\\ 
Energy & Rapidity &\multicolumn{2}{|c|}{Gamma } & \multicolumn{2}{|c|}{shifted-Gamma } &\multicolumn{2}{|c|}{Tsallis }\\ 
 (TeV)& Interval &\multicolumn{2}{|c|}{Distribution} & \multicolumn{2}{|c|}{Distribution} &\multicolumn{2}{|c|}{Distribution}\\ \hline 
 & & & & & & & \\
  & $|y|$ & $\chi^2$/ndf &p value & $\chi^2$/ndf & p value & $\chi^2$/ndf & p value    \\
 & & & & & & & \\\hline

  \multirow{4}{*}{0.9}    & 0.5    &3.21/19	   &1.0000	& 0.98/19	&1.0000 &1.58/17	&1.0000    \\ \cline{2-8}

		&1.0	&54.50/36 	&0.0247	& 33.66/36	&0.5804 &43.64/34	&0.1244\\ \cline{2-8}
		&1.5	&48.49/48  	&0.4531 & 35.41/48	&0.9113 &41.32/46	&0.6683\\ \cline{2-8}
		&2.0	&38.11/58 	&0.9798 & 31.21/58	&0.9985	&33.06/56	&0.9938\\ \cline{2-8}
		&2.4	&52.93/64 	&0.8368 & 44.02/64	&0.9733	&46.82/62	&0.9240\\\hline
 &         &  			  & & &   & &       \\	
  \multirow{4}{*}{2.36}   &0.5	&7.41/19	&0.9917	&5.94/19 &	1.0000   &6.60/17	&0.9882    \\ \cline{2-8}

		&1.0	&67.79/36	&0.0011	&50.59/36 &	0.0541   &59.99/34	&0.0039\\\cline{2-8}
		&1.5	&30.10/46	&0.9662	&25.24/46 &	0.9945   &27.28/44 	&0.9774\\\cline{2-8}
		&2.0	&46.44/56	&0.8162	&45.25/56 &	0.8473   &45.31/54	&0.7941\\\cline{2-8}
		&2.4	&44.91/65	&0.9729	&44.98/65 &	0.9778   &40.94/63	&0.9859\\\hline
 &         &  			  & & &     & &     \\
 					
\multirow{4}{*}{7.00} 	&0.5	&101.40/37	&0.0001	& 75.86/37 &	0.0002	&91.74/35	&0.0001\\\cline{2-8}
		&1.0	&183.71/66	&0.0001	& 149.11/66 &	0.0001		&170.93/64	&0.0001\\\cline{2-8}
		&1.5	&34.93/68	&0.9997	& 35.89/68 &	0.9995 	&35.38/66	&0.9991\\\cline{2-8}
		&2.0	&40.41/86	&1.0000	&  44.27/86 &	0.9999	&41.69/84	&1.0000\\\cline{2-8}
		&2.4	&47.91/99	&1.0000	& 	54.67/99 &	0.9999	&49.96/97	&1.0000\\\hline

\end{tabular}
\caption{$\chi^{2}/ndf$ comparison for the fits with three different distributions for all rapidity windows for the $pp$ data.} 
\end{table*}
\begin{table*}[t]
\begin{tabular}{|c|c|c|c|c|c|c|c|c|c|}
\hline
 &\multicolumn{2}{|c|}{}& \multicolumn{2}{|c|}{} &\multicolumn{4}{|c|}{}   \\
 
 Energy & \multicolumn{2}{|c|}{Gamma distribution} & \multicolumn{2}{|c|}{shifted-Gamma distribution} &\multicolumn{4}{|c|}{Tsallis distribution}\\ \hline
  	    &	          &			 &	  &	  &   &   &  &  \\ 
  	
  	(TeV)  & $\alpha$  & $\beta$ &$\alpha'$ & $\beta'$ &  $nV$ & $nv_{0}$ & K & q \\ 
  	 	
  	 	  &	          &			 &	  &	  &   &   &  &  \\\hline 
  	 
0.9	&1.386 $\pm$ 0.057	&0.211 $\pm$ 0.005 &  1.782 $\pm$ 0.098 &	0.227 $\pm$ 0.006& 2.214 $\pm$ 0.054	&0.249 $\pm$ 0.054	&1.476 $\pm$ 0.056	&1.055 $\pm$ 0.036\\\hline

2.36	&1.204 $\pm$ 0.054	&0.137 $\pm$ 0.003&1.373 $\pm$ 0.093 &	0.142$\pm$ 0.005	& 2.602 $\pm$ 0.106	&0.282 $\pm$ 0.028	&1.241 $\pm$ 0.063	&1.128 $\pm$ 0.007\\\hline

7.00	&1.288 $\pm$ 0.037	&0.096 $\pm$ 0.001	& 1.438 $\pm$ 0.048 &	0.099 $\pm$ 0.001 & 3.611 $\pm$ 0.110	&0.177 $\pm$ 0.093	&1.326 $\pm$ 0.033	&1.242 $\pm$ 0.092\\\hline
\end{tabular}
 
\caption{Fit Parameters with the distributions for charged particle multiplicity spectra in $|y| < 2.4 $ and $P_{T} > $500 MeV of the $pp$ data.} 
\end{table*}

\begin{table*}[t]
\begin{tabular}{|c|c|c|c|c|c|c|}
\hline

&\multicolumn{2}{|c|}{} & \multicolumn{2}{|c|}{} &\multicolumn{2}{|c|}{}\\ 
Energy  &\multicolumn{2}{|c|}{Gamma distribution} & \multicolumn{2}{|c|}{shifted-Gamma distribution } &\multicolumn{2}{|c|}{Tsallis distribution }\\ 
(TeV)&\multicolumn{2}{|c|}{} & \multicolumn{2}{|c|}{} &\multicolumn{2}{|c|}{}\\\hline 
 & & &  & & & \\
   & $\chi^2$/ndf &p value & $\chi^2$/ndf & p value & $\chi^2$/ndf & p value    \\
 & & &  & & & \\\hline

0.9   &32.84/34	&0.5244	&17.86/34	&0.9896	&21.26/32	&0.9177    \\\hline
2.36  &36.37/36	&0.4514	&33.43/36	&0.5914	&35.61/34	&0.3195\\\hline
7.00  &178.07/75	&0.0001	&157.02/75	&0.0001	&172.01/73	&0.0001\\\hline

\end{tabular}
\caption{$\chi^{2}/ndf$ values obtained with the distributions fits to the charged particle multiplicity spectrum for $|y| < 2.4$  and $P_{T} > $500 MeV in the $pp$ data.} 
\end{table*}

\begin{table*}[t]
\begin{tabular}{|>{\centering}m{0.90cm}|c|c|c|}
\hline
  $|y|$ &  A        &  	B		           \\

        &           &                        \\\hline
 
0.5	&0.851 $\pm$ 0.020	&0.076 $\pm$ 0.003\\\hline
1.0	&0.811 $\pm$ 0.072	&0.076 $\pm$ 0.010\\\hline
1.5	&0.706 $\pm$ 0.106	&0.077 $\pm$ 0.018\\\hline
2.0	&0.660 $\pm$ 0.014	&0.076 $\pm$ 0.003\\\hline
2.4	&0.630 $\pm$ 0.081	&0.075 $\pm$ 0.012\\\hline
  \end{tabular}
\caption{Parameters A and B of the Power law fit between q and c.m.energy for the $pp$ collision data.}  
\end{table*}

\begin{table*}[t]
\begin{tabular}{|>{\centering}m{0.90cm}|c|c|c|c|c|c|c|c|c|}
\hline
 
  	    &	          &			 &	  &	  &   &   &   & \\ 
  	
  	$|y|$ & $P_{T}$ & $\alpha_{soft}$ &$\alpha'_{1}$  & $\beta'_{1}$ &$\alpha'_{2}$ & $\beta'_{2}$ &  $\chi^2$/ndf & p values  \\ 
  	 	 	
  	 	  &	   (MeV)       &			 &	  &	  &   &   &  &  \\\hline 
  	 	  
  	 	    	  	  		\multicolumn{9}{|c|}{}  \\ 
  			 \multicolumn{9}{|c|}{$\sqrt{s}$ = 7~TeV}  \\       
			  \multicolumn{9}{|c|}{}    \\\hline
  	  &	        &			 &	  &	  &   &   &  &  \\
0.5	& $>$ 0 	& 0.81	& 2.981$\pm$  0.221	& 0.260$\pm$  0.010	& 7.443$\pm$  0.505	& 1.291$\pm$  0.091	& 11.42 / 35	& 0.9999\\\hline
 &	        &			 &	  &	  &   &   &  &  \\
1.0	& $>$ 0 	& 0.77	& 3.702$\pm$  0.033	& 0.161$\pm$  0.002	& 6.985$\pm$  0.145	& 0.695$\pm$  0.028	& 57.91 / 64	& 0.6904\\\hline
 &	        &			 &	  &	  &   &   &  &  \\
2.4	& $>$ 500	& 0.64	& 5.302$\pm$  0.101	& 0.141$\pm$  0.003	& 3.235$\pm$  0.030	& 0.576$\pm$  0.023	& 69.67 / 73	&0.5888\\\hline

\end{tabular}
 
\caption{Fit Parameters with the 2-component shifted-Gamma distribution to multiplicity spectra in $pp$ collisions for different rapidity windows and $P_{T}$.} 
\end{table*}


\end{document}